\documentclass[aps,prd,onecolumn,eqsecnum,amsmath,nofootinbib,preprintnumbers]{revtex4}%

\newcounter{example}[section]
\newenvironment{example}[1][]{\refstepcounter{example}\par\medskip
	\noindent \textbf{Example~\theexample. #1} \rmfamily}{\medskip}

\usepackage{amsmath}
\usepackage{color,graphicx,float,subfigure}
\usepackage{amsfonts,amssymb,theorem,mathrsfs,times}
\usepackage{bm}
\usepackage{amsmath}
\newcommand{\dalm}{\kern1pt\vbox{\hrule height 0.9pt\hbox{\vrule width
			0.9pt\hskip 2.5pt\vbox{\vskip 5.5pt}\hskip 3pt\vrule width
			0.3pt}\hrule height 0.3pt}\kern1pt}

\begin{document}
	\preprint{\hfill {\small {USTC-6-26}}}
	\title{The evolutions of the innermost stable circular orbits in dynamical spacetimes}
	
	%
	
\author{ Yong Song\footnote{e-mail
		address: syong@mail.ustc.edu.cn}}

\affiliation{Department of Modern Physics, \\
	University of Science and Technology of China, Hefei, Anhui 230026,
	China}
	


	\date{\today}
	
\begin{abstract}
In this paper, we studied the evolutions of the innermost stable circular orbits (ISCOs) in dynamical spacetimes. At first, we reviewed the method to obtain the ISCO in Schwarzschild spacetime by varying its conserved orbital angular momentum. Then, we demonstrated this method is equivalent to the effective potential method in general static and stationary spacetimes. Unlike the effective potential method, which depends on the presence of the conserved orbital energy, this method requires the existence of conserved orbital angular momentum in spacetime. So it can be easily generalized to the dynamical spacetimes where there exists conserved orbital angular momentum. From this generalization, we studied the evolutions of the ISCOs in Vaidya spacetime, Vaidya-AdS spacetime and the slow rotation limit of Kerr-Vaidya spacetime. The results given by these examples are all reasonable and can be compared with the evolutions of the photon spheres in dynamical spacetimes.
\end{abstract}

\maketitle

\section{Introduction}
Accretion disks are ubiquitous in astronomy, and there are usually accretion phenomena around black holes. Through the study of accretion disks, one can obtain a lot of information about black holes. In 2019, the Event Horizon Telescope took the first image of a black hole at the center of the M87 galaxy~\cite{Akiyama:2019cqa}. In the image, one can see a shadow region which is called the black hole shadow, and the black hole lies in the shadow. One can also see a ring-like structure that corresponds to the accretion disk, and the ISCO plays a vital role in analyzing this image~\cite{Akiyama:2019fyp,Kawashima:2019ljv}.

Up to date, there are many studies based on the effective potential to study ISCO in spacetime. On the one hand, ISCO has many important properties. For example, it is the inner edge of an accretion disk ~\cite{Abramowicz:2011xu}, it is the boundary between the stable orbits and the unstable orbits~\cite{Cardoso:2008bp,Cornish:2003ig}, and the accretion flow changes dramatically across the ISCO in a thin disk~\cite{Reynolds:2007rx,Penna:2010hu,Noble:2010mm,McKinney:2012vh}. On the other hand, ISCO has many applications. Such as, for a rotating black hole, the radius of ISCO is a key fit parameter to measure the spin of the black hole~\cite{McClintock:2011zq,McClintock:2013vwa}, and there are many other studies about the ISCOs in Kerr-like spacetimes~\cite{Bardeen:1972fi,Teo:2020sey,Chaverri-Miranda:2017gxq,Tsupko:2016bpn,Favata:2010ic,Hackmann:2010ir,Zhang:2017nhl,Stuchlik:2003dt,Slany:2013ora,Slany:2020jhs}. In the modified gravitational theories, ISCOs may also exist~\cite{Shaymatov:2020yte}. Also, ISCO may have some applications in AdS/CFT. In recent years, some studies suggest that ISCO should describe field theory long-lived excitations that do not thermalize like typical excitations~\cite{Berenstein:2020vlp}.

Through the effective potential method, one can efficiently study the ISCOs in static and stationary spacetimes. But, this method is not suitable for dynamical spacetimes because the effective potential cannot be defined in dynamical spacetimes. 

However, many studies involve dynamical spacetimes. Such as the formation of a black hole~\cite{OConnor:2010moj,Margalit:2015qza,Camelio:2018gfc,Uchida:2018ago}, the specific angular momentum of ISCO is vital to the formation of the disk around the black hole. Because ISCO is the edge of the accretion disk, knowing the evolution of ISCO helps simulate the formation of the disk. To get the evolution equations of the ISCOs in dynamical spacetimes, we ask the following question: Is there a method equivalent to the effective potential method in static and stationary spacetimes and can be easily generalized to the dynamical spacetimes?

In this paper, we reviewed the method to obtain the ISCO in Schwarzschild spacetime by varying its conserved orbital angular momentum. We then demonstrated this method is equivalent to the effective potential method in general static and stationary spacetimes. To illustrate this equivalence further, We studied the ISCOs in general static spherically symmetric spacetimes and Kerr spacetime. The results of ISCOs in these spacetimes are all consistent with the previous results. We then generalized this method into the dynamical spacetimes where there exists the conserved orbital angular momentum. From the generalization, we studied the ISCOs in Vaidya spacetime, Vaidya-AdS spacetime and the slow rotation limit of Kerr-Vaidya spacetime. The results given by these examples are reasonable and can be compared with the evolutions of the photon spheres in dynamical spacetimes~\cite{Mishra:2019trb}. So, we believe that this generalization is reliable. This method only requires the conserved orbital angular momentum in spacetime, so it may have a more widespread application than the effective potential method which depends on the conserved orbital energy in spacetime. As long as there is conserved orbital angular momentum in spacetime, ISCO can be obtained by using this method.

This paper is organized as follows: In Section \ref{section2}, we will study the ISCO in Schwarzschild spacetime and obtain two important properties of ISCO. In section \ref{section3}, We will demonstrate that there is a method equivalent to the effective potential method to study ISCOs in static and stationary spacetimes and use some examples to verify this equivalence. In section \ref{section4}, We will generalize this method to dynamical spacetimes and use some examples to illustrate the reliability of this generalization. Section \ref{conclusion} is devoted to the conclusion and discussion.

Convention of this paper: We choose the system of geometrized unit, i.e., set $G=c=1$. Also, we set the mass of the free point particle $m=1$ and use $M$ to denote the mass of a black hole. The abstract index formalism has been used to clarify some formulas or calculations~\cite{Wald:1984}. A quantity with a lower script ``o" represents the quantity associated with a circular orbit and a lower script ``isco" represents the quantity associated with an ISCO.

\section{ISCO in Schwarzschild spacetime}\label{section2}
In this section, we will review the two methods to get the ISCO in Schwarzschild spacetime. From the second method, we will obtain two crucial properties of ISCO in Schwarzschild spacetime, and these two properties are essential to generalize the second method to dynamical spacetimes.

The metric in $\{t,r,\theta,\phi\}$ coordinates of the $4$-dimensional Schwarzschild spacetime can be written as
\begin{eqnarray}
\label{Schmetric}
ds^2=-\bigg(1-\frac{2M}{r}\bigg)dt^2+\bigg(1-\frac{2M}{r}\bigg)^{-1}dr^2+r^2(d\theta^2+\sin^2\theta d\phi^2)\;.
\end{eqnarray}
At first, we review the method to find the ISCO by using the effective potential method. In Schwarzschild spacetime, the spherical symmetry allows us to choose the equatorial plane, i.e., $\theta=\frac{\pi}{2}$. For a timelike geodesic, the effective potential in the equatorial plane can be defined as~\cite{Wald:1984}
\begin{eqnarray}
V_l(r)=\frac{1}{2}-\frac{M}{r}+\frac{l^2}{2r^2}-\frac{Ml^2}{r^3}\;,
\end{eqnarray}
where $l$ is the conserved orbital angular momentum. For a cirular orbit, one have $V_{l_o}'(r_o)\equiv\left.\partial V_{l}(r)/\partial r\right|_{r_{o},l_o}=0$. Then, we get the following equation
\begin{eqnarray}
V_{l_o}^\prime(r_o)=\frac{M}{r_o^2}-\frac{l_o^2}{r_o^3}+3\frac{Ml_o^2}{r_o^4}=0\;.
\end{eqnarray}
Solving the above equation, we get the orbital angular momentum of a circular orbit as follows
\begin{eqnarray}
\label{lo}
l_o=\sqrt{\frac{Mr_o^2}{r_o-3M}}\;,
\end{eqnarray}
To get the ISCO, one can reqiure that
$V_{l_{isco}}^{\prime\prime}(r_{isco})=0$~\cite{Berenstein:2020vlp}, i.e.,
\begin{eqnarray}
V_{l_{isco}}^{\prime\prime}(r_{isco})=-2\frac{M}{r_{isco}^3}+3\frac{l_{isco}^2}{r_{isco}^4}-\frac{12Ml_{isco}^2}{r_{isco}^5}=\frac{M(r_{isco}-6M)}{(r_{isco}-3M)r_{isco}^2}=0\;,
\end{eqnarray}
where we have restricted eq.(\ref{lo}) in an ISCO. Then we can get the location and the orbital angular momentum of the ISCO in Schwarzschild spacetime as 
\begin{eqnarray}
r_{isco}=6M,\qquad l_{isco}=3\sqrt{2}M\;.
\end{eqnarray}

Below, we will use the second method by analyzing the geodesic equations and varying the conserved orbital angular momentum of the spacetime to obtain the above results. Consider a timelike geodesic in the equatorial plane, its normalized 4-velocity can be expressed as 
\begin{eqnarray}
u^a=\frac{dx^\mu(\tau)}{d\tau}\bigg(\frac{\partial}{\partial x^\mu}\bigg)^a=\frac{dt(\tau)}{d\tau}\bigg(\frac{\partial}{\partial t}\bigg)^a+\frac{dr(\tau)}{d\tau}\bigg(\frac{\partial}{\partial r}\bigg)^a+\frac{d\phi(\tau)}{d\tau}\bigg(\frac{\partial}{\partial \phi}\bigg)^a\;,
\end{eqnarray}
where $\tau$ is its proper time. From the normalized condition of the 4-velocity, i.e., $u^au_a=-1$, we have
\begin{eqnarray}
\label{Schuaua}
-\bigg(1-\frac{2M}{r}\bigg)\bigg(\frac{dt}{d\tau}\bigg)^2+\bigg(1-\frac{2M}{r}\bigg)^{-1}\bigg(\frac{dr}{d\tau}\bigg)^2+r^2\bigg(\frac{d\phi}{d\tau}\bigg)^2=-1.
\end{eqnarray}
The geodesic equation relates to the $r$ coordinate of eq.(\ref{Schmetric}) can be expressed as
\begin{eqnarray}
\label{Schgeodesicr}
\frac{d^2r}{d\tau^2}+\frac{M(r-2M)}{r^3}\bigg(\frac{dt}{d\tau}\bigg)^2-\frac{M}{r(r-2M)}\bigg(\frac{dr}{d\tau}\bigg)^2-(r-2M)\bigg(\frac{d\phi}{d\tau}\bigg)^2=0.
\end{eqnarray}
Considering the trajectory of the timelike geodesic is circular, we can set $r=r_{o}=\mathrm{constant}$, i.e., $dr_{o}/d\tau=0\;,d^2r_{o}/d\tau^2=0$. Then eq.(\ref{Schuaua}) and (\ref{Schgeodesicr}) can be simplified to
\begin{eqnarray}
-\bigg(1-\frac{2M}{r_{o}}\bigg)\bigg(\frac{dt}{d\tau}\bigg)^2+r_{o}^2\bigg(\frac{d\phi}{d\tau}\bigg)^2&=&-1\;,\\
\frac{M}{r_{o}^3}\bigg(\frac{dt}{d\tau}\bigg)^2-\bigg(\frac{d\phi}{d\tau}\bigg)^2&=&0\;.
\end{eqnarray}
Combining the above two equations, we get the following results
\begin{eqnarray}
&&\bigg(\frac{dt}{d\tau}\bigg)^2=\frac{r_o}{r_o-3M}\;,\\
&&\bigg(\frac{d\phi}{d\tau}\bigg)^2=\frac{M}{r^2_{o}(r_{o}-3M)}.
\end{eqnarray}
Notice that the conserved orbital angular momentum of a circular orbit in Schwarzschild spacetime can be defined as
\begin{eqnarray}
\label{Schl}
l_o\equiv\bigg(r_{o}^2\frac{d\phi}{d\tau}\bigg)=\sqrt{\frac{Mr_{o}^2}{r_{o}-3M}}\;.
\end{eqnarray}
Using the well-known result that ISCO has a minimal angular momentum among all circular orbits in Schwarzchild spacetime~\cite{Carroll:1997ar}, i.e., it satisfies
\begin{eqnarray}
\label{Schdl}
\frac{\delta l_o}{\delta r_{o}}=0\;.
\end{eqnarray}
Then, combining eq.(\ref{Schl}) and (\ref{Schdl}), we get the equation of the ISCO as follows
\begin{eqnarray}
\label{Schisco}
-\frac{\sqrt{M}(r_{isco}-6M)}{2(r_{isco}-3M)^{3/2}}=0\;.
\end{eqnarray}
From eq.(\ref{Schl}) and (\ref{Schisco}), we obtain the location and the conserved orbital angular momentum of the ISCO in Schwarzschild spacetime as
\begin{eqnarray}
	r_{isco}=6M\;,\qquad l_{isco}=3\sqrt{2}M\;,
\end{eqnarray}
which are consistent with the previous results. From the above analysis, we get two critical properties of the ISCO in Schwarzschild spacetime:
\begin{itemize}
	\item[(1).] For a general circular orbit, it does not evolve in time, i.e., 
	\begin{eqnarray}
	\label{property1}
	dr_{o}/d\tau=d^2r_{o}/d\tau^2=0\;.
	\end{eqnarray}
	\item[(2).] For a family of circular orbits, ISCO has a minimal orbital angular momentum, i.e.,
	\begin{eqnarray}
	\label{property2}
	\delta l_o/\delta r_{o}=\delta l_o^2/\delta r_{o}=0\;,
	\end{eqnarray}
where $l_o$ should be regarded as a function of $r_o$\footnote{For a circular orbit, $l_o\ne 0$. Sometimes, it is more convenient to use the expression of $l_o^2$ to get the equation of ISCO. If the solution of eq.(\ref{property2}) is single-valued, it is an ISCO. If the solution of eq.(\ref{property2}) is double-valued, such as Schwarzschild-dS spacetime, Kerr-dS spacetime and so on~\cite{Stuchlik:1999qk,Stuchlik:2003dt,Slany:2020jhs,Stuchlik:2020rls}, the one with $\delta^2 l_o/\delta r_o^2>0$ is ISCO, and the one with $\delta^2 l_o/\delta r_o^2<0$ is OSCO (outermost stable circular orbit)~\cite{Boonserm:2019nqq,Berry:2020ntz}, and one can easily check that this is correct in Schwarzschild-dS spacetime~\cite{Stuchlik:1999qk}. On the other hand, the ISCO also has minimal orbital energy among the circular orbits, and  it is a standard way to identify the location of the ISCO by finding the minimum of the orbital energy~\cite{Hod:2014tpa,Favata:2010ic,Favata:2010yd,Buonanno:2002ft}.}.
\end{itemize}
From the above analysis, we realize that these two methods may have some connection. In the next section, we will demonstrate that these two methods actually equivalent under certain conditions. We will then generalize the second method to study the evolutions of the ISCOs in dynamical spacetimes.


\section{ISCOs in static and stationary spacetimes}\label{section3}
In the general static and stationary spacetimes, eq.(\ref{property1}) is obviously valid. Below we will demonstrate that eq.(\ref{property2}) is also valid in some conditions. 

In the general static and stationary spacetimes, suppose one can define the effective potential as $V_l(r)$, where $l$ is the conserved orbital angular momentum. Consider a free point particle, and for a given circular orbit, one always has the following relation 
\begin{eqnarray}
	\label{V1}
	V_{l_o}^\prime(r_o)=0\;,
\end{eqnarray}
and for this circular orbit, $l_o$ is a constant. Considering a family of circular orbits and varying eq.(\ref{V1}), one can get the following equation,
\begin{eqnarray}
	0=\frac{\delta V_{l_o}^\prime(r_o)}{\delta r_o}=V_{l_o}^{\prime\prime}(r_o)+\frac{\partial V_{l_o}^{\prime}(r_o)}{\partial l_o}\frac{\delta l_o}{\delta r_o}\;.
\end{eqnarray}
Here, $l_o$ should regard as a function of $r_o$. Then, one have the following relation
\begin{eqnarray}
\label{Vlgeneral}
V_{l_o}^{\prime\prime}(r_o)=-\frac{\partial V_{l_o}^\prime(r_o)}{\partial l_o}\frac{\delta l_o}{\delta r_o}\;,
\end{eqnarray}
where we have assumed that $\left.\partial V_{l_o}^{\prime}(r_o)/\partial l_o\right|_{r_{isco},l_{isco}}\ne0$. In general, this assumption can be satisfied. So, for an ISCO, the condition $V_{l_{isco}}^{\prime\prime}(r_{isco})=0$ is equivalent to $\delta l_o/\delta r_o=0$\footnote{Similar arguments have been made in~\cite{Berry:2020ntz}. On the other hand, it is equivalent to identifying the ISCO by finding the minimum of the orbital energy and finding the minimum of the orbital angular momentum in static or stationary spacetime~\cite{Damour:2000we}. Suppose there exist conserved orbital angular momentum and conserved orbital energy in the static or stationary spacetime. Considering a free point particle, for a circular orbit, one can always define
\begin{eqnarray}
\label{eo}
e_o\equiv V_{l_o}(r_o)\;.
\end{eqnarray}
where $e_o$ is the orbital energy of the circular orbit. For a family of circular orbits, we consider the variation of eq.(\ref{eo}), i.e.,
\begin{eqnarray}
\frac{\delta e_o}{\delta r_o}=\frac{\partial V_{l_o}(r_o)}{\partial r_o}+\frac{\partial V_{l_o}(r_o)}{\partial l_o}\frac{\delta l_o}{\delta r_o}\;.
\end{eqnarray}
Notice that $\partial V_{l_o}(r_o)/\partial r_o=0$ for a circular orbit, and $\partial V_{l_o}(r_o)/\partial l_o\ne 0$ in general, so the equation $\delta e_o/\delta r_o=0$ is equivalent to $\delta l_o/\delta r_o=0$.}, and the stable circular orbits should satisfy the condition that $\delta l_o/\delta r_o\ge 0$.

Below, we will study the ISCOs in general static spherically symmetric spacetimes and Kerr spacetime to illustrate this equivalence, and one can easily check that eq.(\ref{Vlgeneral}) holds in Schwarzschild spacetime.
\subsection{ISCOs in general static spherically symmetric spacetimes}
The metric of the $(d+1)$-dimensional static spherically symmetric spacetimes in general can be written as
\begin{eqnarray}
	\label{GSmetric}
	ds^2=-f(r)dt^2+g(r)dr^2+r^2d\Omega_{d-1}^2\;,
\end{eqnarray}
where $d\Omega_{d-1}^2$ is the line element of the unit $S^{d-1}$. Similarly, considering a timelike geodesic on the equatorial plane, and from the normalized condition of the 4-velocity, we have
\begin{eqnarray}
	\label{GSuaua1}
	-f\bigg(\frac{dt}{d\tau}\bigg)^2+g\bigg(\frac{dr}{d\tau}\bigg)^2+r^2\bigg(\frac{d\phi}{d\tau}\bigg)^2=-1\;.
\end{eqnarray}
The geodesic equation relates to the $r$ coordinate of eq.(\ref{GSmetric}) can be written as follows
\begin{eqnarray}
	\label{GSGeodesicr}
	&&\frac{d^2r}{d\tau^2}+\frac{f^\prime}{2g}\bigg(\frac{dt}{d\tau}\bigg)^2+\frac{g^\prime}{2g}\bigg(\frac{dr}{d\tau}\bigg)^2-\frac{r}{g}\bigg(\frac{d\phi}{d\tau}\bigg)^2=0\;.
\end{eqnarray}
where a prime denotes a derivative with respect to areal radius $r$. Considering eq.(\ref{property1}), then eqs.(\ref{GSuaua1}) and (\ref{GSGeodesicr}) become to
\begin{eqnarray}
	&&-f(r_o)\bigg(\frac{dt}{d\tau}\bigg)^2+r_{o}^2\bigg(\frac{d\phi}{d\tau}\bigg)^2=-1\;,\\
	&&\frac{f^\prime(r_o)}{2g(r_o)}\bigg(\frac{dt}{d\tau}\bigg)^2-\frac{r_{o}}{g(r_o)}\bigg(\frac{d\phi}{d\tau}\bigg)^2=0\;.
\end{eqnarray}
Combining the above two equations, we get the following equations
\begin{eqnarray}
&&\bigg(\frac{dt}{d\tau}\bigg)^2=\frac{2}{2f(r_o)-r_of^\prime(r_o)}\;,\\
&&\bigg(\frac{d\phi}{d\tau}\bigg)^2=\frac{f^{\prime}(r_o)}{2r_{o}f(r_o)-r^2_of^\prime(r_o)}\;.
\end{eqnarray}
The conserved orbital angular momentum in a circular orbit of the general static spherically symmetric spacetimes can be defined as
\begin{eqnarray}
	\label{staticl}
	l_o\equiv r^2_{o}\frac{d\phi}{d\tau}=\sqrt{\frac{r^3_{o}f^{\prime}(r_o)}{2f(r_o)-f^\prime(r_o) r_o}}\;,
\end{eqnarray}
and from eq.(\ref{property2}), i.e.,
\begin{eqnarray}
	\frac{\delta l_o}{\delta r_o}=0\;,
\end{eqnarray}
we get the equation of the location of the ISCO as follows
\begin{eqnarray}
\label{staticisco}
\frac{2\sqrt{r_{isco}}[r_{isco}f(r_{isco})f^{\prime\prime}(r_{isco})+3f(r_{isco})f^\prime(r_{isco})-2r_{isco}f^{\prime 2}(r_{isco})]}{f^{\prime}(r_o)[2f(r_{isco})-r_{isoc}f^\prime(r_{isco})]^{3/2}}=0\;,
\end{eqnarray}
and the nontrivial part of the above equation is consistent with the nontrivial part of~\cite{Cardoso:2008bp}. 

Let us check whether the above result and the result given by the effective potential method satisfy eq.(\ref{Vlgeneral}). The effective potential in our case should be defined as
\begin{eqnarray}
	V_l(r)=f(r)\bigg(1+\frac{l^2}{r^2}\bigg)\;.
\end{eqnarray}
For a circular orbit
\begin{eqnarray}
V^{\prime}_{l_o}(r_o)=f^{\prime}(r_o)\bigg(1+\frac{l_o^2}{r_o^2}\bigg)-2f(r_o)\frac{l_o^2}{r_o^3}\;,
\end{eqnarray}
and we can get
\begin{eqnarray}
\label{staticV2}
V^{\prime\prime}_{l_{isco}}(r_{isco})=\frac{2r_{isco}f(r_{isco})f^{\prime\prime}(r_{isco})+6f(r_{isco})f^\prime(r_{isco})-4r_{isco}f^{\prime 2}(r_{isco})}{2r_{isco}f(r_{isco})-r^2_{isoc}f^\prime(r_{isco})}\;.
\end{eqnarray}
Then, one can easily check that the difference of eq.(\ref{staticisco}) and (\ref{staticV2}) is exactly $-\partial V_{l_o}^{\prime}(r_o)/\partial l_o$.

As an example, we consider the $(d+1)$-dimensional Schwarzschild-AdS spacetime. The metric of the $(d+1)$-dimensional SAdS in global coordinate can be expressed as
\begin{eqnarray}
	ds^2=-\bigg(1+\frac{r^2}{L^2}-\frac{2M}{r^{d-2}}\bigg)dt^2+\bigg(1+\frac{r^2}{L^2}-\frac{2M}{r^{d-2}}\bigg)^{-1}dr^2+r^2d\Omega_{d-1}^2\;,
\end{eqnarray}
where $L$ is the AdS radius. From eq.(\ref{staticl}), we get the conserved orbital angular momentum in SAdS spacetime as
\begin{eqnarray}
	\label{SAdSl}
	l_o^2=\frac{r_{o}^4}{L^2}\frac{r_{o}^d+(d-2)M L^2}{r_{o}^d-dMr_o^2}\;.
\end{eqnarray}
In order to get the ISCO and for simplification, we use the following condition
\begin{eqnarray}
	\label{SAdSl2}
	\frac{\delta l_o^2}{\delta r_o}=0\;.
\end{eqnarray}
From eq.(\ref{SAdSl2}), we get the equation of ISCO as
\begin{eqnarray}
\label{SAdSisco}
\frac{-2(d-2)dL^2M^2r_{isco}^5+4r_{isco}^{3+2d}-Mr_{isco}^{3+d}\bigg[(d-4)(d-2)L^2+d(d+2)r_{isco}^2\bigg]}{L^2(dMr_{isco}^2-r_{isco}^d)^2}=0\;,
\end{eqnarray}
and the nontrivial part is consistent with the eq.(21) in~\cite{Berenstein:2020vlp} after aligned the equation of $l^2$.


\subsection{ISCO in Kerr spacetime}
Because of the complexity of the calculations in general stationary spacetimes, in this subsection, we will use Kerr spacetime as an example to illustrate the validity of the equivalence in stationary spacetimes.
 
The metric of $4$-dimensional Kerr spacetime in Boyer-Lindquist coordinates can be written as
\begin{eqnarray}
\label{kerr}
ds^2=&-&(1-2Mr/\Sigma)dt^2-(4Mar\sin^2\theta/\Sigma)dtd\phi\nonumber\\
&+&(\Sigma/\Delta)dr^2+\Sigma d\theta^2+(r^2+a^2+2Ma^2r\sin^2\theta/\Sigma)\sin^2\theta d\phi^2\;,
\end{eqnarray}
where $a$ is the angular momentum per unit mass of the black hole ($0\le a\le M$), and the functions $\Delta\;,\Sigma$ are defined as
\begin{eqnarray}
\Delta&\equiv&r^2-2Mr+a^2\;,\\
\Sigma&\equiv&r^2+a^2\cos^2\theta\;.
\end{eqnarray} 
At first, we consider the situation on the equatorial plane, and eq.(\ref{kerr}) can be simplified as
\begin{eqnarray}
ds^2=-\bigg(1-\frac{2M}{r}\bigg)dt^2-\frac{4Ma}{r}dtd\phi+\frac{r^2}{r^2-2Mr+a^2}dr^2+\bigg(r^2+a^2+\frac{2Ma^2}{r}\bigg)d\phi^2\;,
\end{eqnarray}
The condition of the normalization 4-velocity can be expressed as
\begin{eqnarray}
\label{kerruaua}
-\bigg(1-\frac{2M}{r}\bigg)\bigg(\frac{dt}{d\tau}\bigg)^2-\frac{4Ma}{r}\frac{dt}{d\tau}\frac{d\phi}{d\tau}+\frac{r^2}{r^2-2Mr+a^2}\bigg(\frac{dr}{d\tau}\bigg)^2+\bigg(r^2+a^2+\frac{2Ma^2}{r}\bigg)\bigg(\frac{d\phi}{d\tau}\bigg)^2=-1\;.
\end{eqnarray}
The Lagrangian of a particle motion can be written as
\begin{eqnarray}
&&\mathcal{L}=-\frac{1}{2}\bigg(1-\frac{2M}{r}\bigg)\bigg(\frac{dt}{d\tau}\bigg)^2-\frac{2Ma}{r}\frac{dt}{d\tau}\frac{d\phi}{d\tau}\nonumber\\
&&+\frac{1}{2}\frac{r^2}{r^2-2Mr+a^2}\bigg(\frac{dr}{d\tau}\bigg)^2+\frac{1}{2}\bigg(r^2+a^2+\frac{2Ma^2}{r}\bigg)\bigg(\frac{d\phi}{d\tau}\bigg)^2\;.
\end{eqnarray}
From the Euler-Lagrange equation
\begin{eqnarray}
\frac{d}{d\tau}\bigg(\frac{\partial\mathcal{L}}{\partial (dx^\mu/d\tau)}\bigg)=\frac{\partial\mathcal{L}}{\partial x^\mu}\;,
\end{eqnarray}
we get the equation of motion in $r$ direction as
\begin{eqnarray}
\label{kerrr2}
&&\frac{r^2}{r^2-2Mr+a^2}\frac{d^2r}{d\tau^2}+\frac{r(r^2-2Mr+a^2)-r^2(r-M)}{(r^2-2Mr+a^2)^2}\bigg(\frac{dr}{d\tau}\bigg)^2\nonumber\\
&&=-\frac{M}{r^2}\bigg(\frac{dt}{d\tau}\bigg)^2+\frac{2Ma}{r^2}\frac{dt}{d\tau}\frac{d\phi}{d\tau}+(r-\frac{Ma^2}{r^2})\bigg(\frac{d\phi}{d\tau}\bigg)^2\;.
\end{eqnarray}
Considering eq.(\ref{property1}), then eqs.(\ref{kerruaua}) and (\ref{kerrr2}) can be simplified as
\begin{eqnarray}
&&-\bigg(1-\frac{2M}{r_o}\bigg)\bigg(\frac{dt}{d\tau}\bigg)^2-\frac{4Ma}{r_o}\frac{dt}{d\tau}\frac{d\phi}{d\tau}+\bigg(r_o^2+a^2+\frac{2Ma^2}{r_o}\bigg)\bigg(\frac{d\phi}{d\tau}\bigg)^2=-1\;,\\
&&-\frac{M}{r_o^2}\bigg(\frac{dt}{d\tau}\bigg)^2+\frac{2Ma}{r_o^2}\frac{dt}{d\tau}\frac{d\phi}{d\tau}+\bigg(r_o-\frac{Ma^2}{r_o^2}\bigg)\bigg(\frac{d\phi}{d\tau}\bigg)^2=0\;.
\end{eqnarray}
Solving the above equations, we get
\begin{eqnarray}
&&\frac{dt}{d\tau}=\pm\frac{r_o^{3/2}-aM^{1/2}}{[-2aM^{1/2}r_o^{3/2}+r_o^2(r_o-3M)]^{1/2}}\;,\\
&&\frac{d\phi}{d\tau}=\mp\frac{M^{1/2}}{[-2aM^{1/2}r_o^{3/2}+r_o^2(r_o-3M)]^{1/2}}\;,
\end{eqnarray}
or
\begin{eqnarray}
&&\frac{dt}{d\tau}=\pm\frac{r_o^{3/2}+aM^{1/2}}{[-2aM^{1/2}r_o^{3/2}+r_o^2(r_o-3M)]^{1/2}}\;,\\
&&\frac{d\phi}{d\tau}=\pm\frac{M^{1/2}}{[2aM^{1/2}r_o^{3/2}+r_o^2(r_o-3M)]^{1/2}}\;.
\end{eqnarray}
The conserved orbital angular momentum in Kerr spacetime on the equatorial plane can be defined as
\begin{eqnarray}
l_o\equiv p_\phi=\frac{\partial\mathcal{L}}{\partial(d\phi/d\tau)}&=&-\frac{2Ma}{r_o}\bigg(\frac{dt}{d\tau}\bigg)+\bigg(r_o^2+a^2+\frac{2Ma^2}{r_o}\bigg)\bigg(\frac{d\phi}{d\tau}\bigg)\nonumber\\
&=&\frac{\pm M^{1/2}(r_o^2\mp 2aM^{1/2}r_o^{1/2}+a^2)}{r_o^{3/4}(r_o^{3/2}-3Mr_o^{1/2}\pm 2aM^{1/2})^{1/2}}\;.
\end{eqnarray}
From eq.(\ref{property2}), i.e.,
\begin{eqnarray}
\frac{\delta l_o}{\delta r_o}=0\;,
\end{eqnarray}
we get the position of the ISCO as
\begin{eqnarray}
\label{kerrisco}
r_{isco}=3M+\sqrt{a^2+3M^2+A}\pm\sqrt{2(a^2+3M^2)-A+\frac{16a^2M}{(a^2+3M^2+A)^{1/2}}}
\end{eqnarray}
where
\begin{eqnarray}
A=(M+a)^{2/3}(M-a)^{1/3}(3M-a)+(M-a)^{2/3}(a+M)^{1/3}(a+3M)\;,
\end{eqnarray}
where ``$-$" corresponds to the ``direct" and ``$+$" corresponds to the ``retrograde". This result is the same as
\begin{eqnarray}
r_{ms}&=&M\{3+Z_2\mp[(3-Z_1)(3+Z_1+2Z_2)]^{1/2}\}\;,\\
Z_1&\equiv& 1+(1-a^2/M^2)^{1/3}[(1+a/M)^{1/3}+(1-a/M)^{1/3}]\;,\\
Z_2&\equiv&(3a^2/M^2+Z_1^2)^{1/2}\;,
\end{eqnarray}
in~\cite{Bardeen:1972} after some calculations, and $r_{ms}$ has the same meaning of $r_{isco}$.

For the case of a circular orbit, which is not confined to the equatorial plane~\cite{Teo:2020sey,Syunyaev:1986zz}, the conserved orbital angular momentum  can be expressed as
\begin{eqnarray}
l_o=-\frac{2M a r_o^3+(r_o^2+a^2)(a Q\mp\sqrt{\Gamma})}{r_o^2\sqrt{r_o^3(r_o-3M)-2a(aQ\mp\sqrt{\Gamma})}}\;,
\end{eqnarray}
where $Q$ is the Carter’s constant and
\begin{eqnarray}
	\Gamma\equiv Mr_o^5-Q(r_o-3M)r_o^3+a^2Q^2\;.
\end{eqnarray}
By using the following condition
\begin{eqnarray}
	\frac{\delta l_o^2}{\delta r_o}=0\;,
\end{eqnarray}
we get the equation of the ISCO as
\begin{eqnarray}
\label{Qisco}
Q_{isco}=\frac{Mr_{isco}^3(a^2M+3a^2r_{isco}-6M^2r_{isco}+3Mr_{isco}^2-r_{isco}^3)+3M^{3/2}r_{isco}^{5/2}\Delta_{isco}^{3/2}}{4a^2r_{isco}(3M^2-3Mr_{isco}+r_{isco}^2)-4a^4M}
\end{eqnarray}
Where $\Delta_{isco}=r_{isco}^2-2Mr_{isco}+a^2$ and $Q_{isco}$ is the Carter’s constant in ISCO. Eq.(\ref{Qisco}) is the same as the result in~\cite{Teo:2020sey} which is expressed as
\begin{eqnarray}
Q_{ms}=-\frac{Mr^{5/2}[(\sqrt{\Delta}-2\sqrt{Mr})^2-4a^2]}{4a^2(r^{3/2}-M\sqrt{r}-\sqrt{M\Delta})}
\end{eqnarray}
after some calculations. Here $Q_{isco}$ and $Q_{ms}$ have the same meaning.


\section{The evolutions of the ISCOs in dynamical spacetimes}\label{section4}
In general dynamical spacetimes, property (1) does not hold anymore. Enlightened by~\cite{Mishra:2019trb}, we assume
\begin{eqnarray}
\label{Gdr}
dr_{o}(t)=\frac{\partial r_{o}(t)}{\partial t}dt=\dot{r}_o(t) dt\;,
\end{eqnarray}
where $t$ is the coordinate time and a dot stands for the derivative with respect to this coordinate time. As for property (2), we generalize it to the following equation
\begin{eqnarray}
\frac{\delta l_o}{\delta r_o(t)}=\frac{\delta l_o^2}{\delta r_o(t)}=0\;.
\end{eqnarray}
Here, $l_o$ should regard as a function of $r_o(t)$. Because of the conservation of the orbital angular momentum, the above equation means that the evolved ISCO has a minimal orbital angular momentum among the evolved circular orbits\footnote{Because the orbital angular momentum of the spacetime is conserved, so it may easy to know whether the orbital angular momentum exists a minimal value or not. For example, the general spherically symmetric spacetime can be written as eq.(\ref{Gmetric}). Let the parameter $t=t_0=\mathrm{constant}$, i.e.,
\begin{eqnarray}
	ds^2=-f(t_0,r)dt^2+g(t_0,r)dr^2+r^2d\Omega_{d-1}^2\;,
\end{eqnarray}
and this spacetime is a static spherically symmetric spacetime. Supposing this spacetime has a minimal orbital angular momentum among the circular orbits, then the corresponding general spherically symmetric spacetime will have a minimal orbital angular momentum among the evolved circular orbits.}.
\subsection{general spherically symmetric spacetime}
The metric of the general spherically symmetric spacetime in $(d+1)$-dimensional spacetime can be written as
\begin{eqnarray}
	\label{Gmetric}
	ds^2=-f(t,r)dt^2+g(t,r)dr^2+r^2d\Omega_{d-1}^2\;.
\end{eqnarray}
where $f$ and $g$ are functions of $\{t,r\}$ coordinates. As before, because of the spherical symmetry of the system, we can consider a timelike geodesic with normalized 4-velocity in the equatorial plane. From the normalized condition of the 4-velocity, we have the following equation
\begin{eqnarray}
	\label{Guaua1}
	-f\bigg(\frac{dt}{d\tau}\bigg)^2+g\bigg(\frac{dr}{d\tau}\bigg)^2+r^2\bigg(\frac{d\phi}{d\tau}\bigg)^2=-1\;,
\end{eqnarray}
where $\tau$ is the proper time of the timelike geodesic. The geodesic equations of eq.(\ref{Gmetric}) are
\begin{eqnarray}
\label{Geodesict1}
\frac{d^2t}{d\tau^2}+\frac{\dot{f}}{2f}\bigg(\frac{dt}{d\tau}\bigg)^2+\frac{f^{\prime}}{f}\frac{dt}{d\tau}\frac{dr}{d\tau}+\frac{\dot{g}}{2f}\bigg(\frac{dr}{d\tau}\bigg)^2&=&0\;,\\
\label{Geodesicr2}
\frac{d^2r}{d\tau^2}+\frac{f^\prime}{2g}\bigg(\frac{dt}{d\tau}\bigg)^2+\frac{\dot{g}}{g}\frac{dt}{d\tau}\frac{dr}{d\tau}+\frac{g^\prime}{2g}\bigg(\frac{dr}{d\tau}\bigg)^2-\frac{r}{g}\bigg(\frac{d\phi}{d\tau}\bigg)^2&=&0\;.
\end{eqnarray}
Now we set $r_o= r_o(t)$ for the evolution of the radius of the circular orbits. From eq.(\ref{Gdr}), we can get the second derivative of $r_o$ with respect to $\tau$ as follows
\begin{eqnarray}
	\label{Gddr}
	\frac{dr_{o}^2}{d\tau^2}=\ddot{r}_{o}\bigg(\frac{dt}{d\tau}\bigg)^2+\dot{r}_{o}\frac{dt^2}{d\tau^2}\;.
\end{eqnarray}
Then put eq.(\ref{Gdr}) and (\ref{Gddr}) into eq.(\ref{Guaua1})-(\ref{Geodesicr2}), we get the following equations\footnote{Here and after in this subsection, $f$ represents $f(r_{o}(t),t)$, $\dot{f}$ represents $\left.\partial f(r,t)/\partial t\right|_{r_{o}(t),t}$ and $f^\prime$ represents $\left.\partial f(r,t)/\partial r\right|_{r_{o}(t),t}$.}
\begin{eqnarray}
	\label{Guaua2}
	-f\bigg(\frac{dt}{d\tau}\bigg)^2+g\dot{r}^2_{o}\bigg(\frac{dt}{d\tau}\bigg)^2+r_{o}^2\bigg(\frac{d\phi}{d\tau}\bigg)^2&=&-1\;,\\
	\label{Geodesict3}
	\frac{d^2t}{d\tau^2}+\frac{\dot{f}}{2f}\bigg(\frac{dt}{d\tau}\bigg)^2+\frac{f^\prime}{f}\dot{r}_{o}\bigg(\frac{dt}{d\tau}\bigg)^2+\frac{\dot{g}}{2f}\dot{r}^2_{o}\bigg(\frac{dt}{d\tau}\bigg)^2&=&0\;,\\
	\label{Geodesicr4}
	\ddot{r}_{o}\bigg(\frac{dt}{d\tau}\bigg)^2+\dot{r}_{o}\frac{d^2t}{d\tau^2}+\frac{f^\prime}{2g}\bigg(\frac{dt}{d\tau}\bigg)^2+\frac{\dot{g}}{g}\dot{r}_{o}\bigg(\frac{dt}{d\tau}\bigg)^2+\frac{g^\prime}{2g}\dot{r}^2_{o}\bigg(\frac{dt}{d\tau}\bigg)^2-\frac{r_{o}}{g}\bigg(\frac{d\phi}{d\tau}\bigg)^2&=&0\;.
\end{eqnarray}
Combining the above equations, we get
\begin{eqnarray}
&&\bigg(\frac{dt}{d\tau}\bigg)^2=\frac{2f}{2f(f-g\dot{r}_o^2)+gr_o\dot{r}_o (\dot{f}+2f^\prime \dot{r}_o+\dot{g}\dot{r}^2_o)-fr_o(\dot{f}+2\dot{g}\dot{r}_o+g^\prime\dot{r}_o^2+2g\ddot{r}_o)}\;,\\
\label{dphi}
&&\bigg(\frac{d\phi}{d\tau}\bigg)^2=\frac{1}{r_{o}}\frac{f(f^\prime+2\dot{g}\dot{r}_{o}+g^\prime\dot{r}^2_{o}+2g\ddot{r}_{o})-g\dot{r}_{o}(\dot{f}+2f^\prime\dot{r}_{o}+\dot{g}\dot{r}^2_{o})}{2f(f-g\dot{r}^2_{o})+gr_{o}\dot{r}_{o}(\dot{f}+2f^\prime \dot{r}_{o}+\dot{g}\dot{r}^2_{o})-fr_{o}(f^\prime+2\dot{g}\dot{r}_{o}+g^\prime\dot{r}^2_{o}+2g\ddot{r}_{o})}\;.
\end{eqnarray}
By the way, the orbital angular frequency associated with a circular orbit in general spherically symmetric spacetime can be obtained as follows
\begin{eqnarray}
	\Omega_o\equiv\frac{d\phi}{dt}=\bigg[\frac{f(f^\prime+2\dot{g}\dot{r}_{o}+g^\prime\dot{r}^2_{o}+2g\ddot{r}_{o})-g\dot{r}_{o}(\dot{f}+2f^\prime\dot{r}_{o}+\dot{g}\dot{r}^2_{o})}{2fr_o}\bigg]^{1/2}\;.
\end{eqnarray}
Also, we can define the conserved orbital angular momentum in general spherically symmetric spacetime as
\begin{eqnarray}
l_o^2\equiv\bigg(r_{o}^2\frac{d\phi}{d\tau}\bigg)^2\;.
\end{eqnarray}
From eq.(\ref{dphi}), we get the conserved orbital angular momentum as 
\begin{eqnarray}
l_o^2=\frac{r^3_{o}\bigg[f(f^\prime+2\dot{g}\dot{r}_{o}+g^\prime\dot{r}^2_{o}+2g\ddot{r}_{o})-g\dot{r}_{o}(\dot{f}+2f^\prime\dot{r}_{o}+\dot{g}\dot{r}^2_{o})\bigg]}{2f(f-g\dot{r}^2_{o})+gr_{o}\dot{r}_{o}(\dot{f}+2f^\prime \dot{r}_{o}+\dot{g}\dot{r}^2_{o})-fr_{o}(f^\prime+2\dot{g}\dot{r}_{o}+g^\prime\dot{r}^2_{o}+2g\ddot{r}_{o})}\;.
\end{eqnarray}
Then, we can obtain the evolution equation of the ISCO from the assumption that the conserved orbital angular momentum has a minimal value at ISCO, i.e., $\delta l_o^2/\delta r_{o}=0$. But this equation is very complicated, and we do not show it here. In a word, we can use the above method to get the evolution equations of the ISCOs in general spherically symmetric spacetimes.

Below, We will use two simple examples to demonstrate the reliability of our method. In the Vaidya case, we get a reasonable evolution curve which is similar to the evolution curve of photon sphere in~\cite{Mishra:2019trb}, and a similar curve is also obtained in Vaidya-AdS4 spacetime.

\begin{example}[Vaidya spacetimes:]
As an example of the method developed above,  we consider a black hole spacetime with accreting null fluid, i.e., the in-going Vaidya spacetime. The metric of the $4$-dimensional Vaidya spacetime in the in-going null coordinate can be written as~\cite{Vaidya:1951zza}
\begin{eqnarray}
\label{Vaidya}
ds^2=-\bigg(1-\frac{2M(v)}{r}\bigg)dv^2+2dvdr+r^2(d\theta^2+\sin^2\theta d\phi^2)\;,
\end{eqnarray}
The above metric is a solution of Einstein gravity with the following energy-momentum tensor,
\begin{eqnarray}
T_{ab}=\frac{\dot{M}(v)}{4\pi r^2}\delta_{va}\delta_{vb}\;,
\end{eqnarray}
where $``\cdot"=\partial/\partial v$ in the Vaidya spacetime. According to the steps in the previous section to find the evolution equation of the ISCO in general spherically symmetric spacetime, we can obtain the evolution equation of the ISCO in Vaidya spacetime. At first, we can get the conserved orbital angular momentum associated with a circular orbit in Vaidya spacetime as
\begin{eqnarray}
	\label{l2}
	l_o^2\equiv\bigg(r_{o}^2\frac{d\phi}{d\tau}\bigg)^2=\frac{2M^2(v)r_{o}^2+M(v)r^3_{o}(3\dot{r}_{o}-1)-r^4_{o}[\dot{M}(v)+r_{o}\ddot{r}_{o}]}{M(v)r_{o}(5-9\dot{r}_{o})-6M^2(v)+r_{o}^2(r_{o}\ddot{r}_{o}+3\dot{r}_{o}-2\dot{r}^2_{o}+\dot{M}(v)-1)}\;,
\end{eqnarray}
where $r_o$ represents $r_o(v)$. Then, the evolution equation of the ISCO can be obtained by using the following condition
\begin{eqnarray}
	\frac{\delta l_o^2}{\delta r_o}=0\;.
\end{eqnarray}
We do not show the result here because the result is very complicated and can only be solved numerically.\footnote{One can use Mathematica to get the evolution equation of the ISCO by solving the second-order Euler-Lagrangian equation, i.e.,
\begin{eqnarray}
\frac{\partial l_o^2}{\partial r_o}-\frac{d}{dv}\bigg(\frac{\partial l_o^2}{\partial \dot{r}_o}\bigg)+\frac{d^2}{dv^2}\bigg(\frac{\partial l_o^2}{\partial \ddot{r}_o}\bigg)=0.
\end{eqnarray}
} To solve it, one must specify the expression of the mass function and give appropriate boundary conditions. Here, we choose the following mass function,
\begin{eqnarray}
\label{massfunction1}
M(v)=\frac{M_0}{2}\bigg(1+\tanh(v)\bigg)\;.
\end{eqnarray}
In the asymptotic future (i.e., $v\to\infty$), Eq.(\ref{massfunction1}) approaches to a constant value $M_0$. And we can impose the future boundary conditions, i.e., $r_{isco}(v\to\infty)=6M_0$ and $\dot{r}_{isco}=\ddot{r}_{isco}=\dddot{r}_{isco}=0$, to obtian the evolution of the ISCO. Below, we will set $M_0=1$, and the evolution of the ISCO in the Vaidya spacetime is shown in fig.(\ref{Vaidya1}).
\begin{figure}[H]
	\centering
	\includegraphics[width=3.5in]{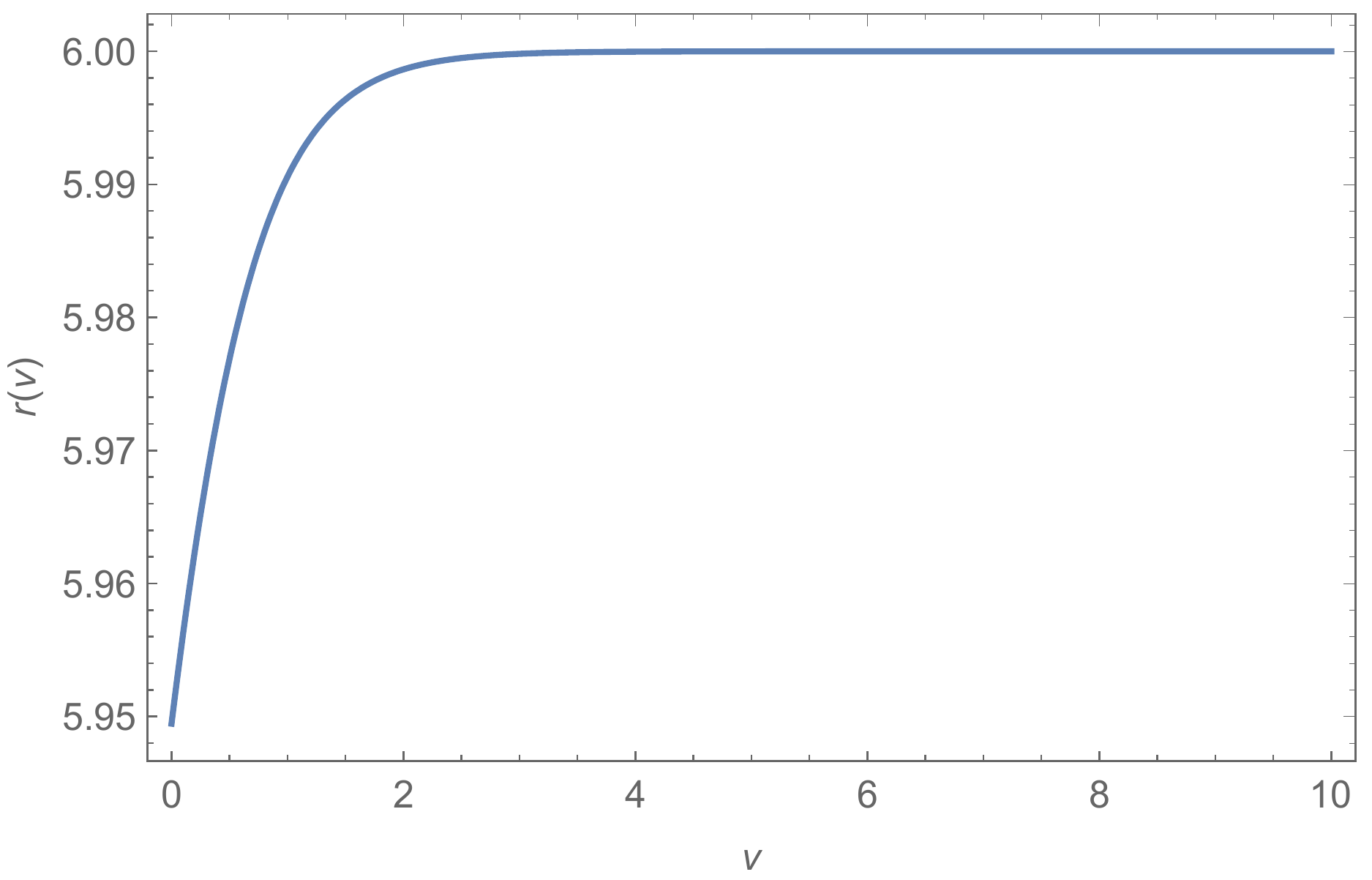}
	\caption{The evolution of the ISCO in the Vaidya spacetime.}
	\label{Vaidya1}
\end{figure}
\end{example}


\begin{example}[Vaidya-AdS spacetimes]
The metric of the $4$-dimensional Vaidya-AdS spacetime can be written as~\cite{Hubeny:2007xt}
\begin{eqnarray}
ds^2=-\bigg(r^2+1-\frac{M(v)}{r}\bigg)dv^2+2dvdr+r^2d\Omega^2_2\;,
\end{eqnarray}
where we have set the AdS radius $L=1$. Following the above procedure, we get the conserved orbital angular momentum associated with a circular orbit in Vaidya-AdS spacetime as
\begin{eqnarray}
l_o^2=r_o^2\frac{2r_o^3(\ddot{r}_o-3r_o\dot{r}_o+r^3+r_o)-r_o(3\dot{r}_o+r_o^2-1)M(v)+r_o^2\dot{M}(v)-M(v)^2}{3M(v)^2+r_o(9\dot{r}_o-3r_o^2-5)M(v)-r_o^2(2r_o\ddot{r}_o-4\dot{r}_o^2+6\dot{r}_o-2r_o^2+\dot{M}(v)-2)}\;.
\end{eqnarray}
Also, the evolution equation of the ISCO can be obtained by using the following assumption
\begin{eqnarray}
	\frac{\delta l_o^2}{\delta r_o}=0\;.
\end{eqnarray}
We do not show the result here because the result is also very complicated and can only be solved numerically. Similar to the Vaidya case, to solve it, one must specify the expression of the mass function and give appropriate boundary conditions. We can get one of the boundary conditions by solving eq.(\ref{SAdSisco}) with $d=3$ numerically, i.e.,
\begin{eqnarray}
r_{isco}\approx 3.76
\end{eqnarray}
in SAdS spacetime, and we have set $M=L=1$. Then, we choose the following mass function (we have already set $M_0=1$),
\begin{eqnarray}
	\label{massfunction2}
	M(v)=\frac{1}{2}\bigg(1+\tanh(v)\bigg)\;.
\end{eqnarray}
Similarly, in the asymptotic future (i.e., $v\to\infty$), we can impose the future boundary conditions, i.e., $r_{isco}(v\to\infty)\approx 3.76$ and $\dot{r}_{isco}=\ddot{r}_{isco}=\dddot{r}_{isco}=0$, to obtian the evolution of the ISCO in Vaidya-AdS4 spacetime. The evolution of the ISCO in the Vaidya-AdS4 spacetime is shown in fig.(\ref{VaidyaAdS}).
\begin{figure}[H]
	\centering
	\includegraphics[width=3.5in]{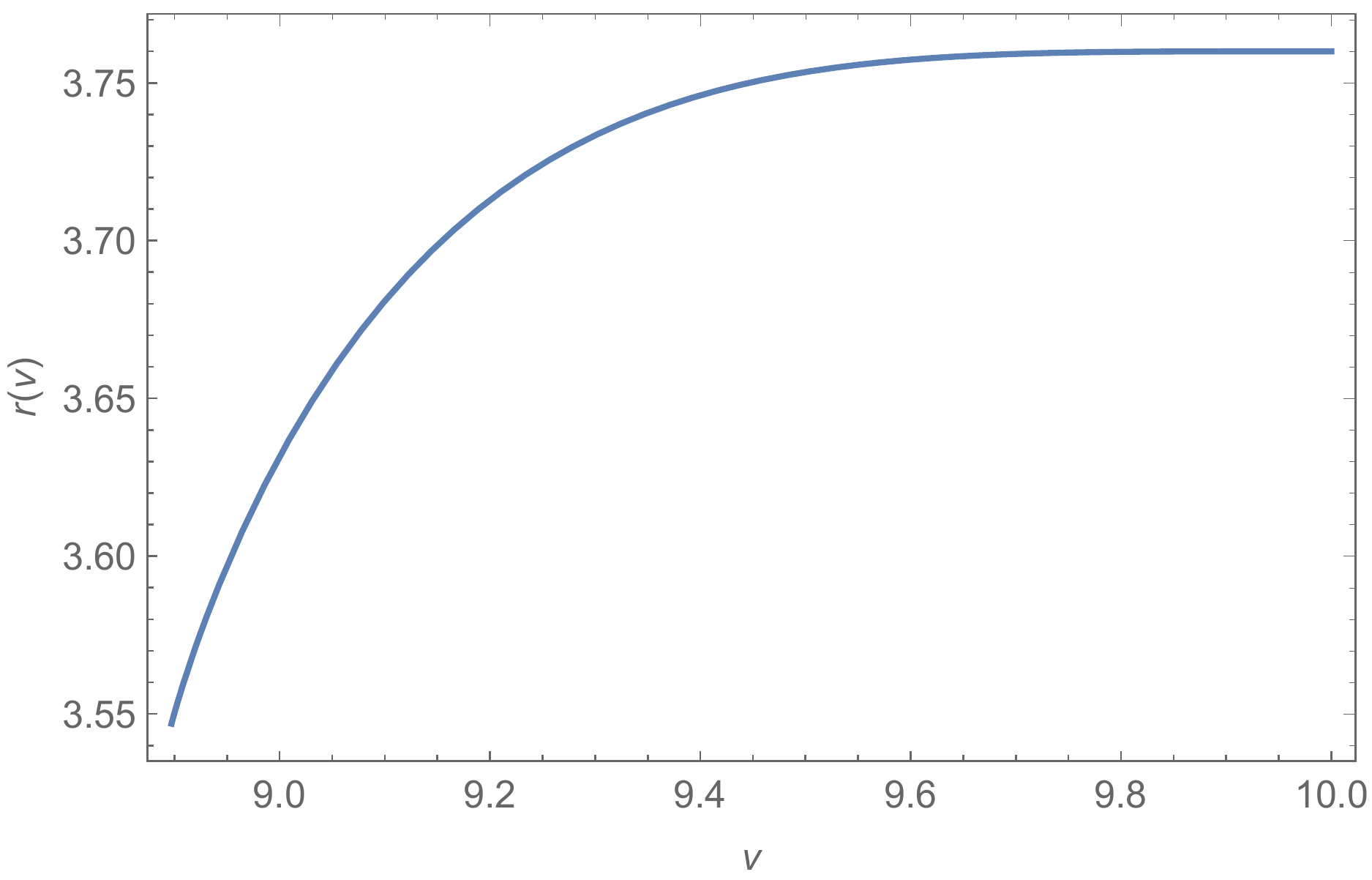}
	\caption{The evolution of the ISCO in the Vaidya-AdS4 spacetime.}
	\label{VaidyaAdS}
\end{figure}
\end{example}

\subsection{ISCO of Kerr-Vaidya spacetime in the slow rotation limit}
In this subsection, we will use the radiating Kerr black hole in the slow rotation limit as an example to show the validity of our method in aspherically symmetric case.

\begin{figure}[H]
	\centering
	\subfigure{
		\begin{minipage}[t]{0.5\linewidth}
			\centering
			\includegraphics[width=3in]{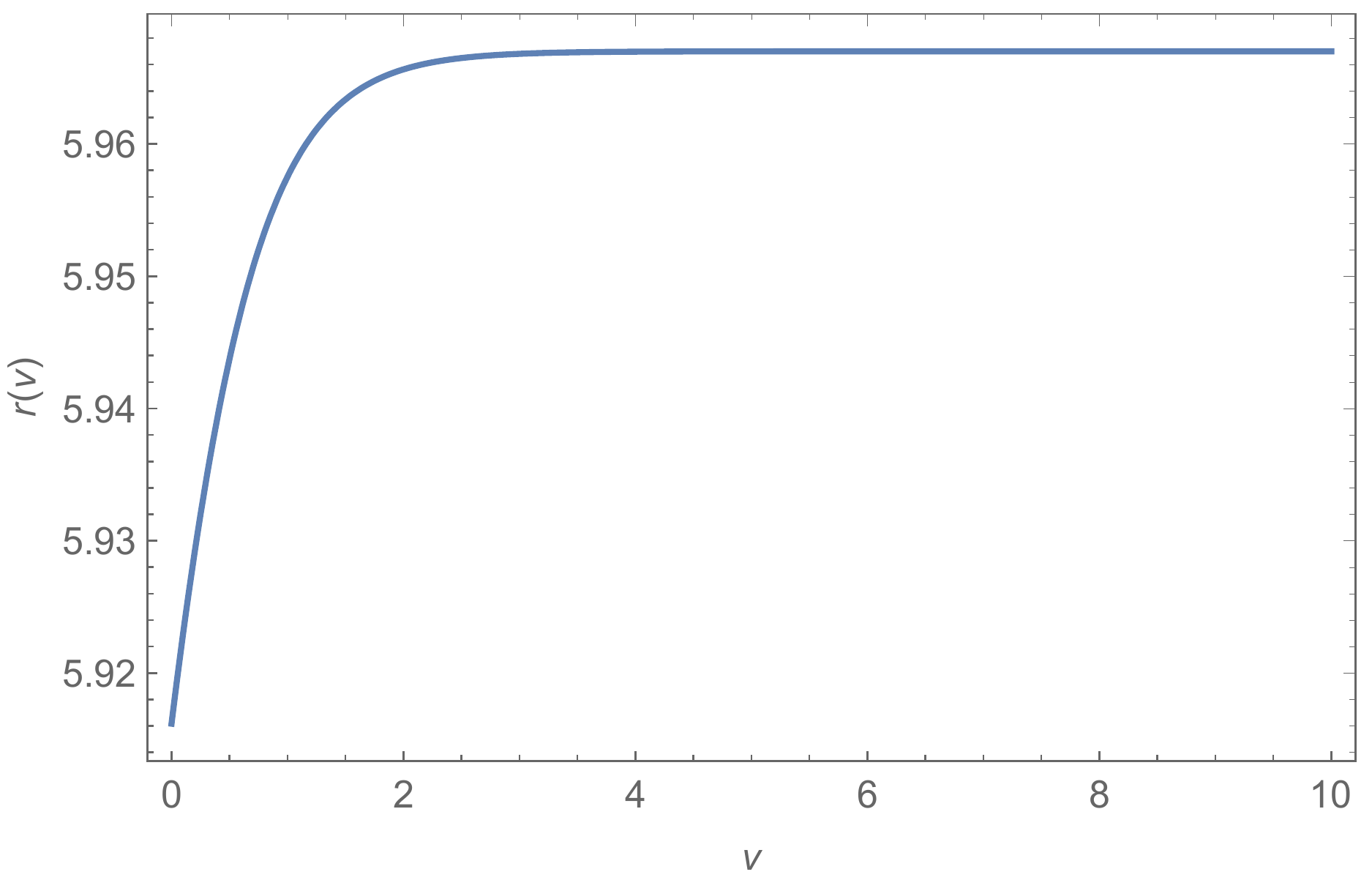}
			\begin{center}
				(a).\, ``direct''
			\end{center}
		\end{minipage}%
	}%
	\subfigure{
		\begin{minipage}[t]{0.5\linewidth}
			\centering
			\includegraphics[width=3in]{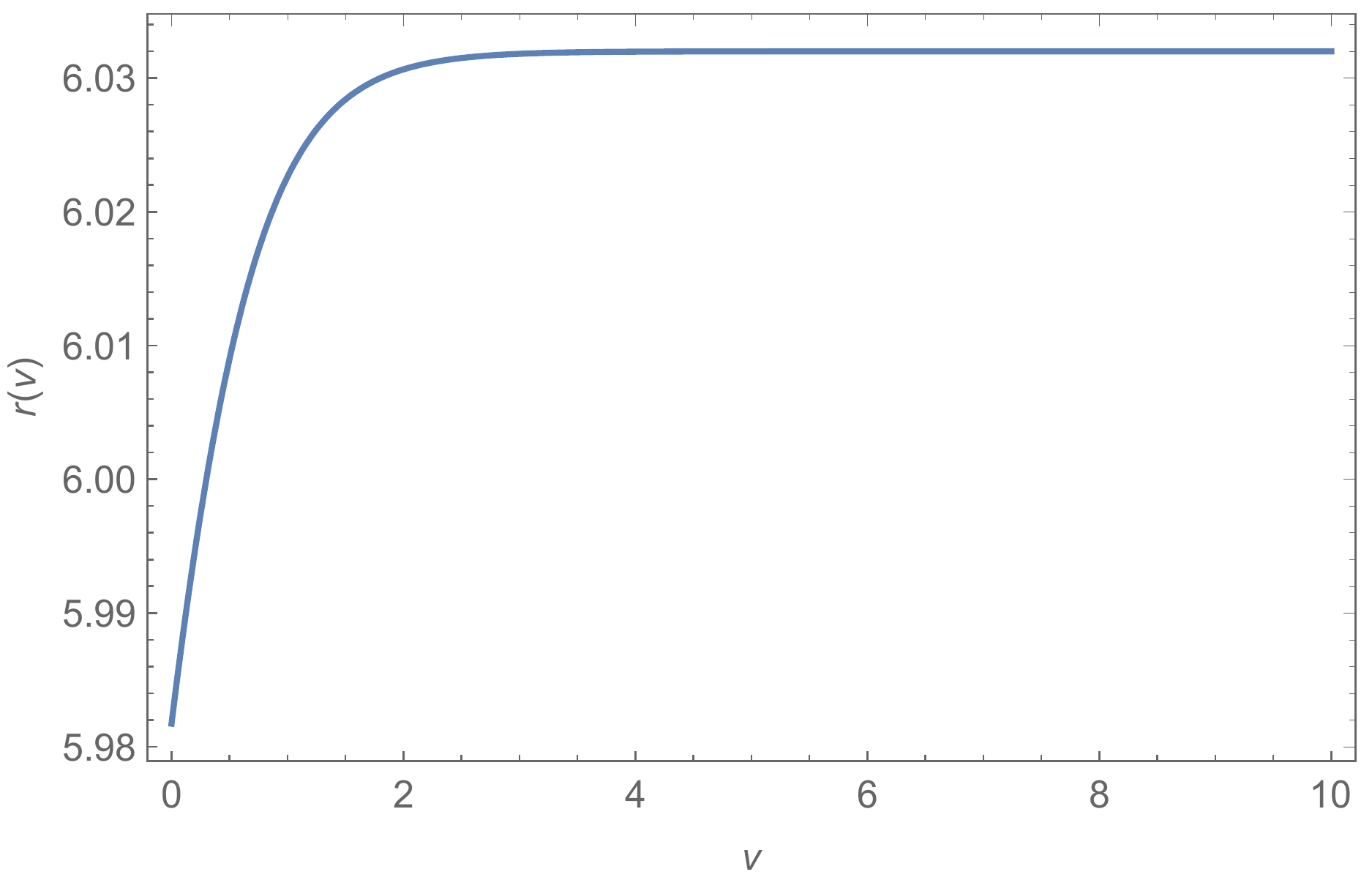}
			\begin{center}
				(b).\,  ``retrograde''
			\end{center}	
		\end{minipage}%
	}%
	\centering
	\caption{The evolution of the ISCO in the slow rotation limit of Kerr-Vaidya spacetime. We have chosen the rotation parameter $a=0.01$.}
	\label{kerrVaidya}
\end{figure}

In the slow rotation limit, the $4$-dimensional Kerr-Vaidya metric on the equatorial plane can be expressed as~\cite{Mishra:2019trb,Murenbeeld:1970aq}
\begin{eqnarray}
ds^2=-\bigg(1-\frac{2M(v)}{r}\bigg)dv^2+2dvdr-2adrd\phi-\frac{4M(v)a}{r}dvd\phi+r^2d\phi^2\;.
\end{eqnarray}
The Lagrangian of a particle motion in this spacetime is
\begin{eqnarray}
\mathcal{L}=-\frac{1}{2}\bigg(1-\frac{2M(v)}{r}\bigg)\bigg(\frac{dv}{d\tau}\bigg)^2+\frac{dv}{d\tau}\frac{dr}{d\tau}-a\frac{dr}{d\tau}\frac{d\phi}{d\tau}-\frac{2M(v)a}{r}\frac{dv}{d\tau}\frac{d\phi}{d\tau}+\frac{r^2}{2}\bigg(\frac{d\phi}{d\tau}\bigg)^2
\end{eqnarray}
Using the normalized condition of the 4-velocity and referencing the formulas in~\cite{Mishra:2019trb}, we can get the following equations for a circular orbit as
\begin{eqnarray}
-\bigg(1-\frac{2M(v)}{r_o}-2\dot{r}_o\bigg)\bigg(\frac{dv}{d\tau}\bigg)^2-\bigg(2a\dot{r}_o+\frac{4aM(v)}{r_o}\bigg)\bigg(\frac{dv}{d\tau}\bigg)\bigg(\frac{d\phi}{d\tau}\bigg)+r_o^2\bigg(\frac{d\phi}{d\tau}\bigg)^2&=&-1\;,\\
\bigg[\ddot{r}_o+\frac{M(v)}{r_o^2}\bigg(1-3\dot{r}_o-\frac{2M(v)}{r_o}\bigg)+\frac{\dot{M}(v)}{r_o}\bigg]\bigg(\frac{dv}{d\tau}\bigg)^2
+\bigg[\frac{2aM(v)}{r^2}\bigg(2\dot{r}_o-1+\frac{2M(v)}{r_o}\bigg)\nonumber\\+\frac{2a\dot{r}_o}{r_o}(1-\dot{r}_o)\bigg]\bigg(\frac{dv}{d\tau}\bigg)\bigg(\frac{d\phi}{d\tau}\bigg)+\bigg(r_o\dot{r}_o-r_o+2M(v)\bigg)\bigg(\frac{d\phi}{d\tau}\bigg)^2&=&0.
\end{eqnarray}
Solving the above two equations, and defining the conserved orbital angular momentum associated with a circular orbit as\footnote{Here we consider $l^2$ instead of $l$ to get a better form and right numerical result, and they are a little different because the higher-order terms that are dropped are different.}
\begin{eqnarray}
l_o^2\equiv\bigg[-\bigg(a\dot{r}_o+\frac{2a M(v)}{r_o}\bigg)\frac{dv}{d\tau}+r_o^2\frac{d\phi}{d\tau}\bigg]^2\;,
\end{eqnarray}
we get
\begin{eqnarray}
&&l_o^2=\frac{2M^2(v)r_{o}^2+M(v)r^3_{o}(3\dot{r}_{o}-1)-r^4_{o}[\dot{M}(v)+r_{o}\ddot{r}_{o}]}{M(v)r_{o}(5-9\dot{r}_{o})-6M^2(v)+r_{o}^2(r_{o}\ddot{r}_{o}+3\dot{r}_{o}-2\dot{r}^2_{o}+\dot{M}(v)-1)}\nonumber\\
&&\pm \frac{6M(v)[2M(v)+r_o(2\dot{r}_o-1)]^2\sqrt{2M(v)+r_o(\dot{r}_o-1)}\sqrt{2M^2(v)r_{o}+M(v)r^2_{o}(3\dot{r}_{o}-1)-r^3_{o}[\dot{M}(v)+r_{o}\ddot{r}_{o}]}}{[M(v)r_{o}(5-9\dot{r}_{o})-6M^2(v)+r_{o}^2(r_{o}\ddot{r}_{o}+3\dot{r}_{o}-2\dot{r}^2_{o}+\dot{M}(v)-1)]^{2}}a\nonumber\\
&&+O(a^2)\;,
\end{eqnarray}
where ``$+$" correspond to the orbital angular momentum associated with ``direct" circular orbit and ``$-$" correspond to the orbital angular momentum associated with ``retrograde" circular orbit. Then, by using the assumption
\begin{eqnarray}
\frac{\delta l_o^2}{\delta r_o}=0,
\end{eqnarray}
we can get the evolution of the ISCO in Kerr-Vaidya spacetime. The explicit form of this evolution equation is very complicated and can only be solved numerically.  Similar to the examlpes of the dynamical Spherically symmetric spacetimes, in order to solve the evolution equation, one must specify the expression of the mass function and give appropriate boundary conditions. We can get one of the boundary conditions by solving the square of eq.(\ref{kerrisco}) with $a=0.01\;, M=1$, i.e.,
\begin{displaymath} 
r_{isco} \approx \left\{ \begin{array}{ll}
	5.967 & \textrm{``direct"}\\
    6.032 & \textrm{``retrograde"}
\end{array} \right. \;,
\end{displaymath}
where we have used the condition of the slow rotation limit in Kerr spacetime. Then, we choose the following mass function (we have already set $M_0=1$),
\begin{eqnarray}
	\label{massfunction3}
	M(v)=\frac{1}{2}\bigg(1+\tanh(v)\bigg)\;.
\end{eqnarray}
Similarly, in the asymptotic future (i.e., $v\to\infty$), we can impose the future boundary conditions, i.e., $r_{isco}(v\to\infty)=5.967$ (direct) and $6.032$ (retrograde), and $\dot{r}_{isco}=\ddot{r}_{isco}=\dddot{r}_{isco}=0$, to obtian the evolution of the ISCO in Kerr-Vaidya spacetime. The evolution of the ISCO of Kerr-Vaidya spacetime in the slow rotation limit is shown in fig.(\ref{kerrVaidya}).

Conclusion of this section: We have generalized the method which is equivalent to the effective potential method in static and stationary spacetimes to general dynamical spacetimes, and use three examples to illustrate the reliability of this generalization. Due to the reasonable results of the evolutions of ISCOs in the examples, we believe this generalization is reliable.

\section{Discussion and conclusion}\label{conclusion}
In this paper, we reviewed the two methods to get the ISCO in Schwarzschild spacetime. We domenstrated the second method is equivalent to the effective potential method in static and stationary spacetimes. We verify this equivalence in general spherically symmetric spacetimes and Kerr spacetime. We then generalized the second method into dynamical spacetimes. From this generalization, we studied the evolutions of the ISCOs in Vaidya spacetime, Vaidya-AdS spacetime, and Kerr spacetime under the limit of slow rotation. These examples are all giving reasonable resluts.

The boundary conditions are essential needed to solve the evolution equations of the ISCOs in dynamical spacetimes. Because the evolution equations are fourth-order equations (we do not show these equations in the paper), so they need four boundary conditions. In general, it is hard to get the appropriate boundary conditions.

From this generalization, one may study the ISCO in more complicated spacetimes as long as there exists a conserved orbital angular momentum. However, the obvious limitation of this method is that it is not suitable for the situation where there is no conserved angular momentum but conserved energy. In this case, the effective potential method may be used to get the ISCO.

\section*{Acknowledgement}
This work was supported in part by the National Natural Science Foundation of China with grants
No.11622543. We would like to thank Li-Ming Cao and Yuxuan Peng for their useful discussions and kindly helps.

\end{document}